\documentclass[pre,twocolumn, showpacs, superscriptaddress]{revtex4}
\usepackage{amsmath}
\usepackage{graphicx}
\usepackage{subfigure}
\usepackage{hyperref}

\def \etal {\emph{et al}}

\def \si {\ensuremath{\sigma_i}}

\def \sr {\ensuremath{\sigma^r}}
\def \sif {\ensuremath{\sigma_i^f}}
\def \sf {\ensuremath{\sigma^f}}
\def \itil {\ensuremath{\tilde{i}}}
\def \metric {\ensuremath{\Omega(t)}}

\begin{document}
\title{A Cellular Automaton Model of Damage}
\author{C. A. Serino}
\email{cserino@physics.bu.edu}
\affiliation{Department of Physics, Boston University, Boston, Massachusetts 02215}
\author{W. Klein}
\affiliation{Department of Physics, Boston University, Boston, Massachusetts 02215}
\affiliation{Center for Computational Science, Boston University, Boston, Massachusetts 02215}
\author{J. B. Rundle}
\affiliation{Department of Physics and Center for Computational Science and Engineering, University of California, Davis, California 95616}
\date{\today}
\begin{abstract}
We investigate the role of equilibrium methods and stress transfer range in describing the process of damage. We find that equilibrium approaches are not applicable to the description of damage and the catastrophic failure mechanism if the stress transfer is short ranged. In the long range limit, equilibrium methods apply only if the healing mechanism associated with ruptured elements is instantaneous. Furthermore we find that the nature of the catastrophic failure depends strongly on the stress transfer range. Long range transfer systems have a failure mechanism that resembles nucleation. In short range stress transfer systems, the catastrophic failure is a continuous process that, in some respects, resembles a critical point.  
\end{abstract}


\maketitle

\section{Introduction\label{sec:intro}}

For reasons of both scientific interest and applications to materials the subject of damage has interested both the physics and material science community for decades. Models of damage such as the fiber bundle model (FBM) \cite{daniels45, peirce26} and the hierarchical model \cite{newetal95} have been used to obtain a greater understanding of the mechanisms of damage and the relationship between damage and phase transitions. \cite{pradhan08, sornette} The treatment of these models has included studies of the effect of various ways the individual fibers break \cite{herrmann1989, hidalgo2002}, investigations of the nature of failure if healing mechanisms are thought to allow the system to remain in equilibrium up to the time of catastrophic failure \cite{robin91} and the effect of inhomogenieties. \cite{herrmann1989, sahimi03} 

Since these models are often treated with equilibrium methods and the time scale of the catastrophic failure event makes its description by equilibrium methods inappropriate, we are left to conisder can the {\it approach} to catastrophic failure be treated with equilibrium methods? Equilibrium methods are often applicable, for example, when catastrophic failure is described by  nucleation, where the probability of the occurrence of a critical droplet can be calculated by assuming a metastable equilibrium state. \cite{langer67,rundlekl89}

In this work we study extensions of the FBM to determine the applicability of equilibrium methods. In addition we probe the nature of the catastrophic failure event and how it is affected by the range of stress transfer. First, we consider the relationship between healing and ergodicity which we accomplish using the Thirumalai - Mountain (TM) metric. \cite{TM90}
This will provide information as to the role equilibrium techniques can play in descriptions of damage. Second, we look at the nature of the critical point that appears to be seen in the global load sharing FBM. \cite{hidalgo2002} This will naturally bring us to a discussion of the role of healing in FBMs.
Finally, we do a careful analysis of the role the stress transfer range has in the nature of the damage process and catastrophic failure.

The structure of the remainder of this paper is as follows: In Section~\ref{sec:model} we introduce the base model and the several variations that we study; in Section~\ref{sec:ergodic} we introduce the TM metric, describe its application to the FBM and our models and discuss the implications of our measurements on the validity of equilibrium descriptions of damage;  \cite{robin91} in Section~\ref{sec:critpt} we investigate the nature of the critical point in the global load sharing FBM; \cite{hidalgo2002} in Section~\ref{sec:range} we investigate the impact the stress transfer range has on the nature of the damage as well as the nature of the catastrophic failure; and in Section~\ref{sec:conc} we present our conclusions.

\section{\label{sec:model}The Model}

We introduce a continuous, cellular automaton (CA) model of damage adapted from the earthquake fault model introduced in 1991 by Olami, Feder, and Christensen.  \cite{OFC92} The Olami-Feder-Christensen (OFC) model is a two-dimensional, CA model motivated by the Burridge--Knopoff spring-block model of earthquake faults. \cite{BK67} The OFC model is equivalent to the CA model proposed by Rundle, Jackson, and Brown (RJB) except that in the latter, there is a natural definition of internal energy, which makes it simpler to identify whether or not the system is in equilibruim. \cite{rundle1977, rundle1991} The evolution of our model is Markovian and described by the following rules. Each site on a lattice (which we take as a $d=2$ square lattice) is assigned a failure threshold $\sigma^{f}_{i}$ and a residual stress $\sigma^{r}_{i}$. For the sake of simplicity, in this work we will take the failure threshold and residual stress to be the same on each site, i.e. $\sigma_i^{r/f} \to \sigma^{r/f}$. If a site's stress reaches or exceeds its failure threshold, the site reduces its stress to the residual stress by dissipating $\alpha\left(\si-\sr\right)$ of its stress(where $0 \leq \alpha < 1$ is a parameter of the model) and passing the remaining fraction of stress (i.e. $(1-\alpha)\left(\si-\sr\right)$) uniformly to the its $q \sim R^{d}$ neighbors. The quantity $q$ can range from nearest neighbor ($q=4$) to ``infinite range '' where $q=N$ is the number of sites in the system. We initialize the system by assigning a random stress satisfying $\sigma^{r} \leq \sigma_{i}< \sigma^{f}$ to each site. Given our initializing procedure, it is clear that at $t=0$ no sites will have $\si\geq\sf$ and hence we must have a procedure for inducing failures. We refer to this process as a plate update. There are several ways to do this but in this paper we consider the so-called zero velocity limit introduced in ref. \cite{OFC92}. According to this procedure, we search the lattice for the site, \itil, that minimizes $\sf - \si$. Next, we add an equal amount of stress to each site such that the stress on \itil\ is now equal to its failure threshold. We then discharge the site per the procedure above and search the lattice to see if the stress added to the neighbors of \itil\ caused any of them to fail. If so, we discharge their stress as above, and if not, we increase the time-step (measured in terms of plate updates) by unity and search the lattice for the next site $\itil'$, which minimizes $\sf - \si$. Note that in this version of the model a site can still receive and hold stress after it fails. We can add noise by resetting sites to a randomized residual stress  and thus instead of the stress on a sight dropping to $\sigma^{r}$ it becomes $\sigma^{r}\pm \eta$. This defines the time evolution of the OFC model. 

Considerable work has been done on this model with noise in the limit that $R\rightarrow \infty$. \cite{rundle1995, ferguson99, klein97} The system has been shown to be in equilibrium and the probability distribution was shown to be Boltzmann. \cite{rundle1995, klein97} To better understand the meaning of the results of our work, it will be useful to discuss the properties of the undamaged model in the $R\rightarrow \infty$ limit presented in references~\cite{rundle1995, ferguson99, klein97}. We begin with  Klein {\it et al} \cite{klein97} where the authors derived a Langevin equation for the time evolution of the stress in the RJB model. In this equation, all lengths can be scaled by $R \propto q^{1/d}$. When this length is scaled out, the noise, assumed to by random Gaussian, must be scaled according to
\begin{equation}
\eta({\vec x},t)\rightarrow \frac{\eta\left(\frac{\vec{x}}{R}, t \right)}{R}\,.
\end{equation}
In the limit $R\rightarrow \infty$, the Langevin equation becomes linear in the stress as all higher order terms are suppressed by powers of $1/R$. This is explained in greater detail in Klein {\it et al}. \cite{klein2000, klein2007} In the linearized Langevin equation, the drift term can be written as the functional derivative of a quadratic action, which guarantees that the probability of a state $\sigma(|{\vec x}|/R)$ approaches a Boltzmann distribution as $t \to \infty$. Additionally, by converting the Langevin equation into a Fokker-Planck equation where the time derivative of the distribution function can be written in terms of the divergence of a probability current, it was shown that the stationary solution to this differential equation causes the current to vanish: a general definition of equilibrium. By calculating the spectrum of eigenvalues of the Fokker-Planck operator, it is scene that the Boltzmann solution is the unique, stable solution to which all initial conditions evolve in the mean-field limit. This was numerically confirmed by Rundle {\it et al} \cite{rundle1995}, in which the authors make measurements of the energy in the system and find its histogram is given by 
\begin{equation}
P(E)\propto g(E)\,e^{-\beta E}\,,
\end{equation}
where $E$ is the energy stored in the springs, $g(E)$ is the density of states which is independent of $\beta$, and $\beta$ is the inverse temperature which is related to the amplitude of the noise by a fluctuation-dissipation relation.

As the system is in equilibrium, it is expected that the dynamics obey some type of detailed balance. We stress that the OFC and RJB models are not in equilibrium for a finite stress transfer range, and so we do not expect detailed balance in the non-mean-field case. We also note that the Langevin/Fokker-Planck treatment is based on coarse graining, which implies that there are coarse grained length and time scales below which the theory does not describe the system. As such, the dynamics of the model described in the beginning of this section do not obey detailed balance as they apply to the cells of the automata which exist on a length scale much, much smaller than the coarse grained length and all of the interactions occur in a single time step which is necessarily much, much smaller than the coarse graining time. Thus, the system can be thought of as in equilibrium, only if it is examined on length scales larger than the coarse graining size and on time scales greater than the coarse graining time and it is at this level that the system obeys a type of detailed balance. In the coarse grained treatment of the RJB and OFC models the authors take the coarse graining length to be the stress transfer range $R$. The coarse graining time is set by the time that the system takes to reach its steady state distribution (i.e. local equilibrium) in the coarse grained volume. This coarse graining time
time goes to infinity as the coarse grained volume diverges ($R\rightarrow \infty$). These details are addressed in full in Ferguson {\it et al}. \cite{ferguson99} Due to these coarse grained length scales, if the system is examined on a microscopic level, it may not appear time reversal invariant and hence a movie of the system run in reverse would look strange. 
However, if observations are only allowed at the coarse grained level, then we would see the stress fluctuating in an apparently random, and time reversal invariant, manner. Similar effects have been seen in references~\cite{Zia, Schmittmann, Praestgaard}.

In the undamaged OFC or RJB model, there is a critical stress, or load, that corresponds to a spinodal, \cite{klein97} that marks the limit of the metastable state and is responsible for the Gutenburg-Richter scaling of event sizes. \cite{klein97} A spinodal is a critical point and the scaling of the avalanche sizes, ``earthquake" magnitudes, or number of sites that fail in a single plate update, is a consequence of fluctuations about the spinodal.\cite{klein2000, klein2007}
The theoretical description of the OFC model in the limit of infinite stress transfer range \cite{klein97} has the same physics as the theoretical description of the TFB model in Selinger \etal. Namely they are both described by a Langevin or Landau-Ginsburg equation with a one component scalar order parameter with the same symmetry. In addition the critical slowing down associated with the spinodal as calculated from the Langevin equation derived in ref.\cite{klein97} has the same critical exponent as that associated with the time to failure in the global stress transfer TFB models studied in ref. \cite{hidalgo2002}. One then expects that the behavior of the two models is the same in this infinite range stress transfer limit. In particular in our model there appears to be a metastable state which ends in a spinodal consistent with the work of Selinger \etal.

However, the class of FBMs treated by Selinger \etal\ and Virgilii \etal\ are unrealistic in that the stress transfer range is not infinite in real materials. In addition healing of the ruptured elements, if it exists, will not be instantaneous. The studies of the OFC model referenced above 
show that the properties of this class of models depends on the stress transfer range. Systems with long but not infinite range interactions 
have pseudo-spinodals, \cite{klein2007, binder1984} metastable states with nucleation, and are in a state of punctuated equilibrium. That is, they appear to be in equilibrium for long periods of time until a large event (``earthquake'') forces the system out of equilibrium. After some relaxation time the system returns to the quasi-equilibrium state and the process repeats. Models with short range stress transfer (nearest neighbor stress transfer for example) show no evidence of being in equilibrium at any time and also show no evidence of a (pseudo) spinodal. \cite{rundle1997, klein2000} In this paper we use the OFC model and variations that we describe herein to study the effect of healing rates, noise, and the stress transfer range.

\subsection{The Base Model}
The simplest version of our modified model is essentially the same as the OFC earthquake fault model with the difference that after a sites fails a given number of times (which we call the site's ``number of lives"), it is considered dead and no longer interacts with the system. Note that this implies that when a site fails within the interaction range of a dead site, the live sites receive more stress in the transfer process than they would if the site were alive. In other words, the stress that would have been passed to the dead site is \emph{not} dissipated, rather, it is shared equally among the remaining live sites within the interaction range. We have also investigated the case in which the stress \emph{is} passed to dead sites and therefore is dissipated. We will discuss the latter case in Section IV. But unless otherwise specified, stress is not transfered to the dead sites.
Given these dynamics, if each site has ten lives, then the evolution of the system when no site has more than nine failures would be identical to the evolution according to the OFC model; however, on the tenth failure, the site dies and it no longer holds, or receives, stress. In order to get rid of transient effects, the system is run without allowing sites to die (hereafter, we call this earthquake mode) for $10^6$ plate updates. After the system reaches a steady state we begin evolving it as a damage model taking note each time a site fails and removing it from the system after a specified number of failures. 

Various changes can be made to the base model to account for different types of materials. 
The failure thresholds can be homogeneous i.e. $\sif = \sf \:\:\forall\:\: i$ to simulate a homogenous or pure material or one could let each site have its own failure threshold to better mimic impurities in a sample or a heterogeneous material. Another way to study the effects of impurities or the general behavior of heterogeneous materials would be let each site have a different number of lives.

\subsection{The Step-down Model}
Even if no stress is added to a system the load bearing sites weaken over time. For example, a system of fibers bearing a constant load will eventually fail. This suggests that the failure thresholds in our model should, themselves, be dynamic and decrease over time. We mimic this behavior by maintaining a fixed total stress on the system and
reducing the failure threshold on a site each time that site fails. This model can be run in several different ways. For example, one could, as in the base model, simply specify the number of times a site must fail before it dies and reduce the failure threshold by a fixed amount each time a site fails. Another method would be to define a critical failure threshold such that a site is dead once its failure threshold drops below this critical value. In addition there is considerable freedom in the method of lowering the failure threshold. The threshold could be reduced by a random amount or a fixed amount. Further, it is known \cite{whiskers}  that micro-cracks can heal on some time scale. In order to capture this phenomena, the failure thresholds could be drawn from a random distribution whose upper bound decreases in time. Thus, when a given site fails, it has some probability of increasing its failure threshold, and some probability of decreasing it, however, by reducing the upper bound of this distribution, one guarantees that, on average, failure thresholds will be reduced. In this paper we will only consider step down models which kill sites when their failure threshold drop below some critical value. Results of simulations of heterogeneous systems will be reported in a future publication.

\subsection{The TFB Model}

Since one of our goals is to understand how damage affects equilibrium states we will also study the thermodynamic fiber bundle model (TFB) model introduced by Selinger \etal. This will serve as a baseline for our studies of the other two classes of models defined above.
To recover the TFB model and the Disordered Thermodynamic Fiber Bundle (DTFB) model of Virgilii \etal\ we must take $R\to\infty$ to ensure global load sharing, and we must not dissipate stress from the system and hence we set $\alpha=0$. In both the TFB and DTFB models, 
the system is described by a Boltzman factor \cite{robin91, DTFB} constructed from the hamiltonian
\begin{equation}
\label{tfb-ham}
\mathcal{H} = \sum_{j}^{N}s_j\left(D_j+\frac{1}{2}\kappa\epsilon^2\right),
\end{equation}
where the $s_j$ equals unity for intact fibers and zero for broken fibers, $D_j$ is the dissociation energy of the $j^{\text{th}}$ fiber, $\kappa$ is the elastic modulus, and $\epsilon$ is the strain on the fiber. \cite{robin91, DTFB} For the TFB model $D_j = D \:\: \forall \:\: j$ whereas for the DTFB model it is fiber dependent \cite{robin91, DTFB}. To make contact between the (D)TFB and our models, we note that all models discussed herein use the results of Brenner \cite{whiskers} to restrict our attention to the Hooke's law regime where $\sigma_i=\kappa\epsilon_i$. In order to simulate these models, we use a Metropolis algorithm. 

\section{Measurements of Ergodicity\label{sec:ergodic}}

As discussed above
some theoretical treatments of damage utilize equilibrium methods \cite{robin91, DTFB} to obtain analytic results for simple models of fracture. The assumption that fracture can be described by an equilibrium theory is often justified by the work of Brenner \cite{whiskers} who performed experiments on iron whiskers and found that the whiskers were, individually, in equilibrium up to the point of failure. It is important to note that Brenner's concept of equilibrium is not necessarily the same as the notion of the equilibrium of statistical ensembles. Brenner finds that the stress-strain curve of {\it each iron whisker} displays no hysteresis \cite{whiskers} up to the point of fracture and thus concludes the fibers are in equilibrium. 
However, this is not a sufficient condition for ergodicity of systems with many, coupled degrees of freedom. Indeed, the process of fracture is an inherently irreversible one and hence any model which truly captures the underlying physics must be non-ergodic on some time scale. The question of interest is the length of that time scale, that is, does the system remain ergodic until the point of catastrophic failure or is the physics of damage essentially non-equilibrium. If it is the latter, then equilibrium methods cannot be applied in analytic work.

In order to test the ergodicity of our model, we measure the stress-fluctuation metric, \metric. This metric was introduced by Thiumalai and Mountain in 1990 \cite{TM90} and adapted to study driven, dissipative systems under stress by Ferguson \etal\ \cite{ferguson99}.
The stress-fluctuation metric is a measure of the difference between the time averaged stress on a site, $\overline \sigma_j(t)$, and the spacial average of the time averaged stress, $\left<\overline{\sigma}(t)\right>$, which approaches the ensemble average for $N\gg 1$. Thus, $\Omega(t)$ is given by
\begin{equation}
\Omega(t) = \frac{1}{N'}\sideset{}{'}\sum_j\left(\overline \sigma_{j}(t) - \left<\overline{\sigma}(t)\right>\right)^2,
\end{equation}
where the sum runs overs the $N'$ \emph{non-failed} sites on the lattice, and the quantities $\overline\sigma_{j}(t)$ and $\left<\overline{\sigma}(t)\right>$ are given by
\begin{equation}
\overline\sigma_{j}(t) = \frac{1}{t}\int_{0}^{t}\text{d}t'\:\sigma_j(t'),
\end{equation}
and
\begin{equation}
\left<\overline{\sigma}(t)\right> = \frac{1}{N'}\sideset{}{'}\sum_j\overline\sigma_{j}(t).
\end{equation}
For effectively ergodic systems, $\Omega(t)\sim 1/t$ and hence plots of $1/\Omega(t)$ versus $t$ will be linear with positive slope.  \cite{TM90}

As mentioned above the OFC model has been exhaustively studied as an earthquake fault model \cite{ferguson99} and we know that the model is ergodic provided the interaction is long range and some noise is introduced into the system. Typically, the noise is added by redrawing the residual stress values from a flat, random distribution of width $\Delta\sigma^r$ and mean $\sigma^r$ each time a site fails. Thus, if we let the long range system evolve for some time in earthquake mode before we let the sites die, we know the system will be ergodic before the first site dies. Therefore we will be able to measure how long the system remains ergodic by studying the time-evolution of the metric. 

The first model we test is the TFB model of Selinger \etal. \cite{robin91} We ran the simulation for $N=256^2=65\,536$ fibers and measured the metric with $\sigma / N=10^{-5}$, $\kappa = 1$, $D = 1$, and $T = 0.5$, following Selinger \etal. \cite{robin91} We choose a fiber at random, switch its state (e.g. broken to intact) and if the energy change is $\Delta E < 0$ we accept the move and if it is $\Delta E > 0$ we accept the move with probability $\exp\left(-\beta \Delta E\right)$. Fibers heal immediately in this model so that as soon as a fiber heals, that fiber supports its equal share of the load (i.e. its elongation and thus stress is the same as the stress on all other intact fibers). The inverse temperature $\beta$ is treated as a parameter in the problem. As one might expect from dynamics that satisfy detailed balance,  the TFB model is ergodic (see Fig. 1) since the inverse metric is a straight line after some transient time. In the work of Selinger \etal\ \cite{robin91} the free energy of the TFB model is shown theoretically to have a metastable and a stable minimum. The failure process is then assumed to be a nucleation event for moderate applied global stress. However, in our simulations of the TFB model with a moderate applied stress the metastable state is has an infinite lifetime. This does not mean that the infinitely long lived state is not meta-stable in the sense that it is a relative minimum of the free energy. Mean-field systems in fact are known to have infinitely long lived metastable states. \cite{klein2007} (It should be noted that the simulations in Selinger \etal\ are not of the TFB model.) Therefore, the TFB with global load sharing cannot capture the physics of catastrophic failure in a fiber bundle model. \cite{note1}

\begin{figure}
\begin{center}
\includegraphics[width = \columnwidth]{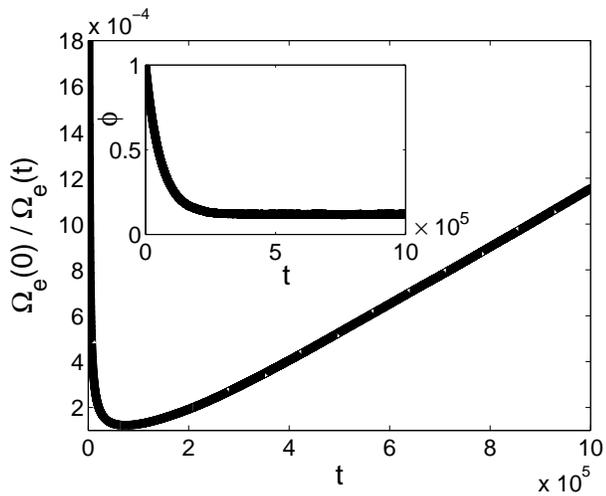}
\caption{The inverse TM  metric and  the order parameter, $\phi=N_{\text{unbroken}}/N$,  (shown in the inset) both as a function of time for the TFB with $N=256^2=65\,536$ fibers, $\sigma / N=10^{-5}$, $\kappa = 1$, $D = 1$, and $T = 0.5$. The metric shows that once the order parameter reaches its metastable value, the system becomes effectively ergodic.}
\label{fig:TFBmetric}
\end{center}
\end{figure}

In addition to the inability of the TFB with global load sharing to actually generate the catastrophic failure mode, it also incorporates the assumptions that the fibers heal instantaneously and the system is in equilibrium. The instantaneous healing assumption is not universally applicable and the question of whether or not systems with individual fibers that have no hysteresis are in equilibrium needs to be tested.  By definition of the model through a Hamiltonian, the states of the system are described by a Boltzmann distribution and the system as a whole is necessarily in equilibrium. Thus, to test these equilibrium assumptions, we need to consider a model where the evolution is described by more microscopic physics. This brings us to our set of models. 

First, we examine the base model we introduced in Section~\ref{sec:model}. We run these systems with large values of the dissipation parameter
$\alpha$ to slow down the failure so that we may record a significant amount of data as cracks appear in the model. We also run the model with three different stress transfer ranges, $R=1$, $R=30$, and with $R$ such that $q=N$, on a square lattice of size $N$ with periodic boundary conditions, and with $N_{\text{lives}}=1$, $\sf = 2.0$,  $\alpha= 0.2$, and $\sr = 1.25 \pm 0.25$, which is the noise as described in our description of the model.

\begin{figure}
\begin{center}
\subfigure[]{\includegraphics[width = \columnwidth]{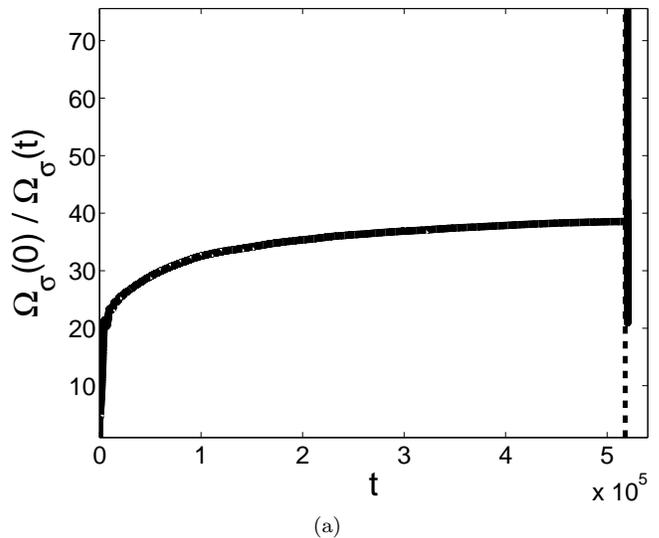}}\hspace{5mm}
\subfigure[]{\includegraphics[width = \columnwidth]{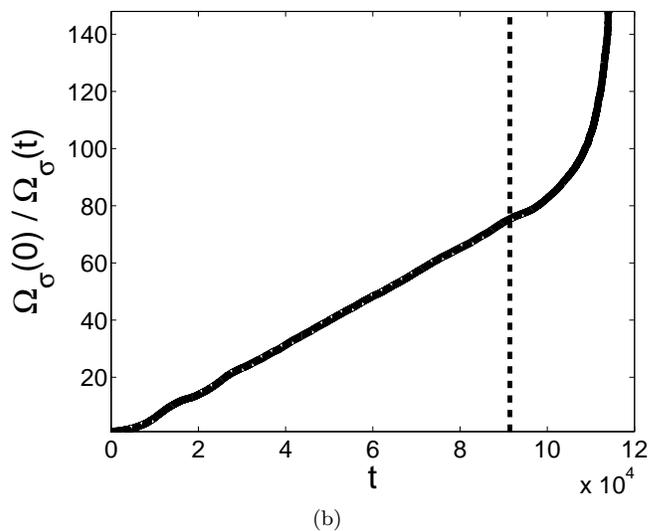}\label{fig:Basemetric30}}\hspace{5mm}
\subfigure[]{\includegraphics[width = \columnwidth]{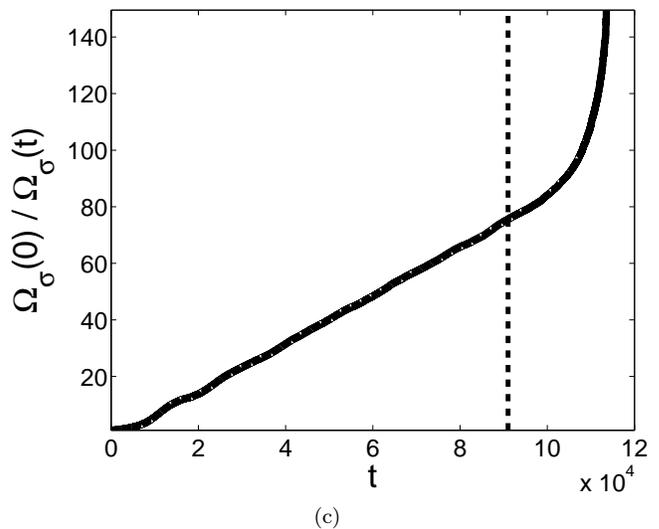}\label{fig:BasemetricInf}}
\caption{The inverse TM metric for the nearest-neighbor (R=1) (a), $R=30$ (b), and mean field ($q=N$) (c) base model simulated on a $d=2$, periodic, square lattice of linear size $L = 100, 512, 768$ for $R = 1, 30,``\infty"$, respectively. The parameters of the model are $\sf = 2.0$, $\sr = 1.25\pm0.25$, and $\alpha= 0.2$. The solid line is the inverse TM metric and the verticle dashed line indicates the time at which the first site dies.}
\label{fig:Basemetric}
\end{center}
\end{figure}

\begin{figure}
\begin{center}
\includegraphics[width = \columnwidth]{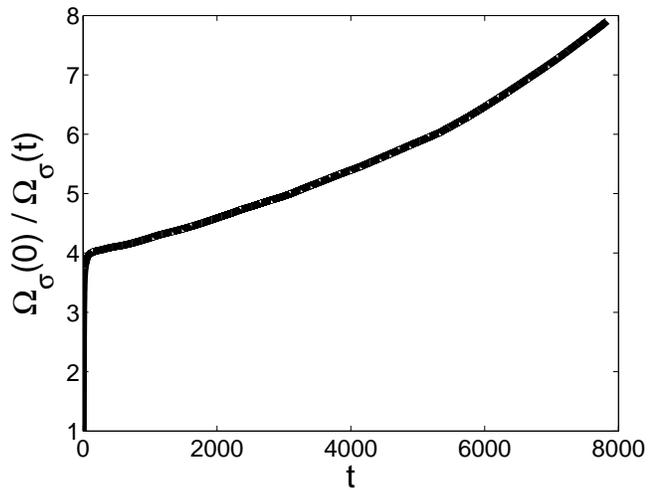}
\caption{The inverse TM metric for the base model when sites are healed after one plate updates. The parameters are given by $R=20$, $L=512$, $\sf = 2$, $\sr=1\pm0.2$, and $\alpha = 0.1$.}
\label{fig:TM_heal}
\end{center}
\end{figure}

\begin{figure}
\begin{center}
\includegraphics[width = \columnwidth]{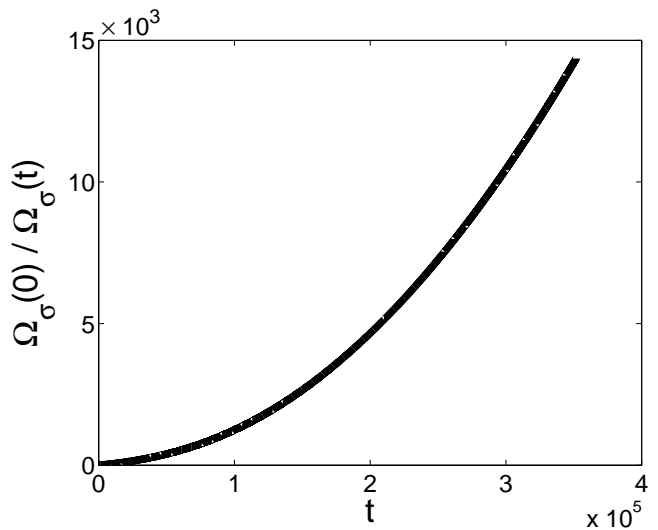}
\caption{The inverse TM metric for the $R=10$ step down model. The model is simulated on a square lattice of linear size $L=256$ with periodic boundary conditions and with $\sigma_f(t=0)=50$, $\sigma_r=0.25\pm0.25$, and $\alpha = 0.05$. We consider a site dead when its failure threshold drops below some predefined value which we take as $\sigma^f_{j}(t) < \sigma^r_j+10^{-5}\Delta$ where $\Delta = \sigma^f(t=0)-\left<\sigma^r\right>=49.75$.}
\label{fig:TM_SD}
\end{center}
\end{figure}

In order that the system be in a steady-state prior to any damage, we let the system run in earthquake mode for $10^6$ plate updates. The metrics for the three interaction ranges are shown in Fig.~\ref{fig:Basemetric}.
The nearest neighbor system is not ergodic even before sites are allowed to die. As we can see from the figure, when sites begin to die the metric shows an even stronger deviation from ergodicity. Systems with longer range interactions show a slightly different behavior. As we know the infinite range OFC model is in equilibrium \cite{klein2000} and a finite but long range interaction exhibits punctuated equilibrium (see the region to the right of the dashed vertical line in Figs.~\ref{fig:Basemetric30} and \ref{fig:BasemetricInf}). When sites begin to die, however, the systems ceases to be in equilibruim (punctuated or otherwise) as is seen in the region after the dashed line in Figs.~\ref{fig:Basemetric30} and \ref{fig:BasemetricInf}. We also measured the TM metric for the base model with healing and long range ($R=20$) stress transfer. In this simulation we allow dead sites to heal after a proscribed number of time steps. Instantaneous healing is simply the OFC model and we know, as mentioned above, that this model is ergodic and incapable of undergoing catastrophic failure. Clearly, as can be seen from the base model, without healing (i.e. the healing time $\to\infty$) the system is not ergodic. In Fig.~\ref{fig:TM_heal} we plot the inverse metric for the base model where the dead sites heal after one plate update. 
As can be seen, the system is clearly not ergodic. Measurements for systems with longer healing times (not shown) are also not ergodic.  Thus, the small change from instantaneous healing (OFC model) to healing after a single plate update (SD model) not only results in a system capable of undergoing catastrophic failure, it also results in a system that is not ergodic on any time interval. Finally, we study the step-down (SD) model. The TM metric for the SD model with $R=10$ is plotted in Fig.~\ref{fig:TM_SD}. 
Note that in Fig.~\ref{fig:TM_SD} the data stops near $t\approx 8\times 10^3$ because the system undergoes catastrophic failure in that time step.

\section{Nature of the Critical Point\label{sec:critpt}}

As we stated in the introduction, the OFC model with long range stress transfer has been shown to have a spinodal critical point. \cite{klein97} It is the spinodal critical point that is responsible for the Gutenburg-Richter scaling in the model. Generally in the neighborhood of the spinodal the nucleation process is not classical. That is, nucleation is not initiated by a compact droplet that has the structure of the stable phase in it's interior. Instead, near the spinodal, the droplet that initiates nucleation is ramified and can be described as a percolation cluster. \cite{klein2007,monette92} As we will see in Section~\ref{sec:range} the catastrophic failure mode in the long range stress transfer base model appears to resemble classical nucleation. This raises the question as to how dying sites affect the spinodal seen in the OFC model. 
To answer that question, we let $\phi \equiv N_{\text{alive}} / N$ parameterize the damage in the system, and we run our model until $\phi = 0.9$ (10\% of the sites die) and then run it in the earthquake mode where $10\%$ of the lattice still consists of dead sites, however, we do not kill any additional sites. We measure the number of clusters ($n_s$) of size $s$, where a cluster is defined as a set of lattice sites that fail as the result of a common ``parent" site having failed. For example, suppose we force a failure per the protocol described in Section~\ref{sec:model} and when the forced site (the so-called ``parent site") passes its stress to its neighbors, three of the neighbors fail. The failed neighbors will, in turn, pass their stress to their neighbors. Let us further suppose that as a result of one of the three neighbors that failed, an additional site fails. Finally, we suppose when this site passes its stress, no more sites fail and thus the event has stopped (i.e. all sites in the lattice have $\sigma_i < \sigma^f$). In this case, all the sites failed as a result of the initial site being forced to fail. These sites, including the ``parent" site, define a cluster, and in this example, the size of the cluster is five. We then run the system in the base model mode until $\phi=0.8$ and repeat the measurement of $n_{s}$ (not shown). We continue this process by similarly decreasing $\phi$. Here we investigate two cases: first we consider the case where the dead sites still receive stress and thus dissipate it from the system and then we consider the case when the dead sites do not receive stress at all. 
\begin{figure}
\begin{center}
\subfigure[]{\includegraphics[width = \columnwidth]{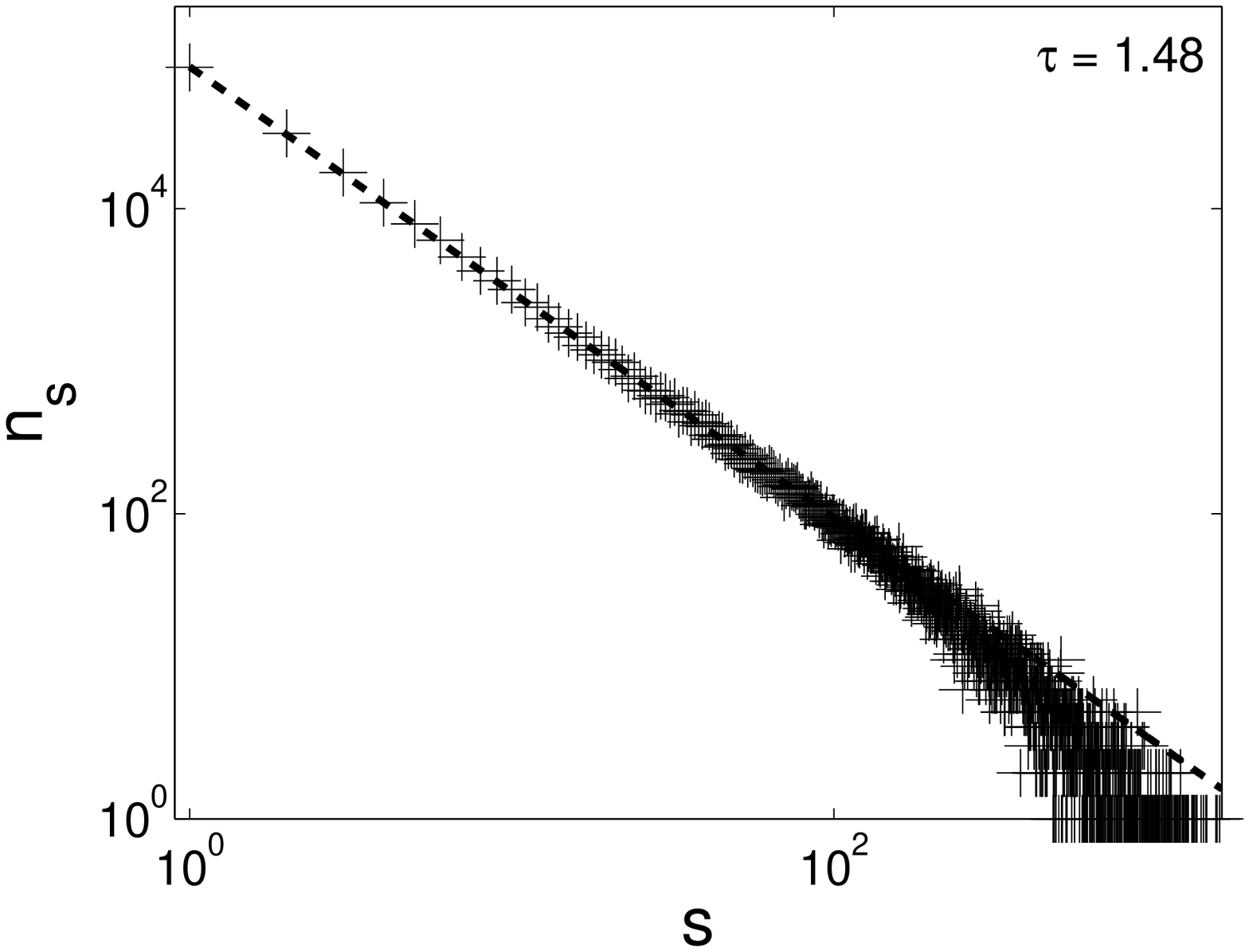}\label{sfig:1}}\hspace{5mm}
\subfigure[]{\includegraphics[width = \columnwidth]{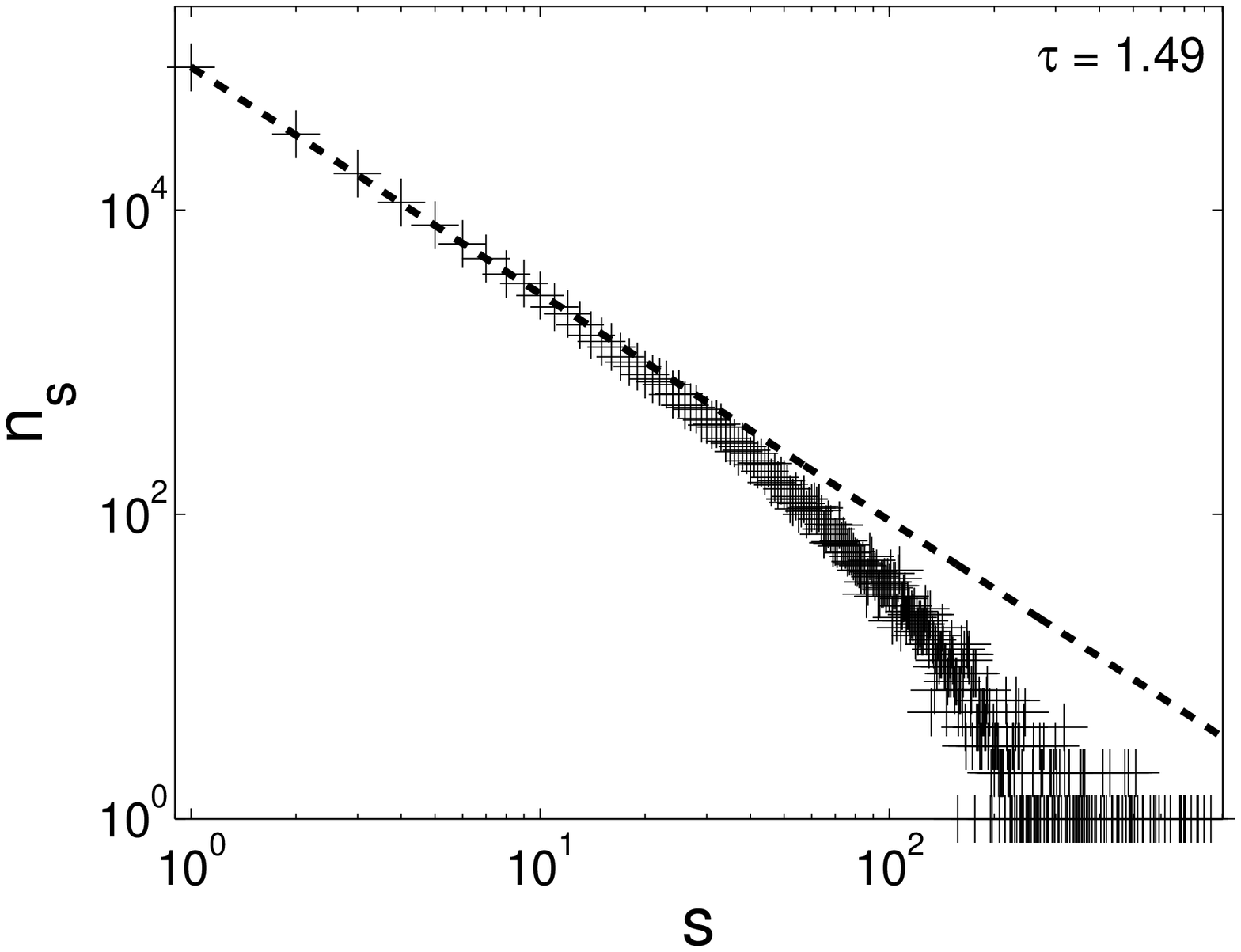}\label{sfig:09}}\hspace{5mm}
\subfigure[]{\includegraphics[width = \columnwidth]{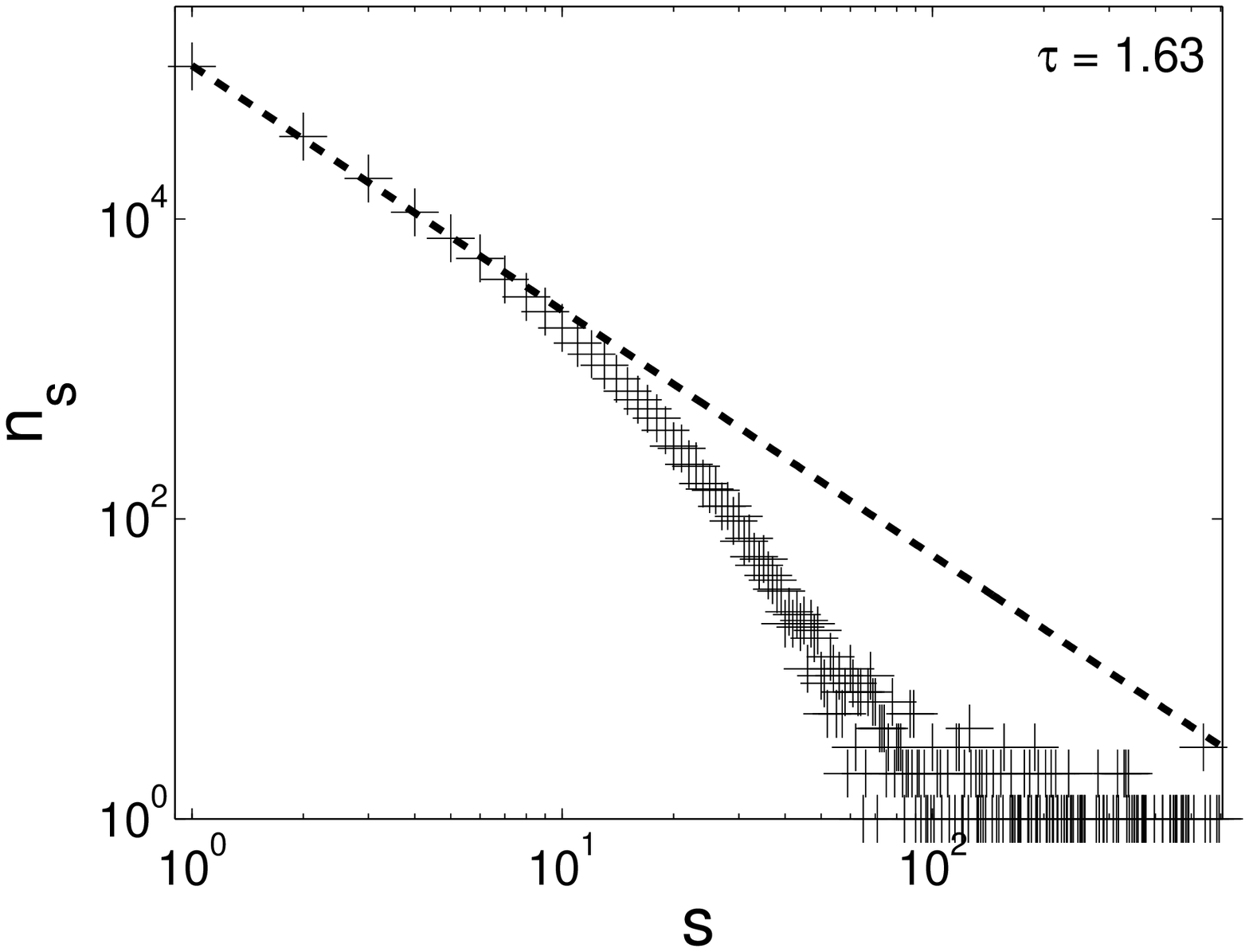}\label{sfig:07}}
\caption{The number of clusters ($n_s$) of size $s$ plotted on a log-log plot for various values of $\phi=N_{\text{alive}}/N$ where stress is passed to all site, alive or dead. For $\phi=1$, we reproduce the known results of the OFC model where $n_s\sim s^{-\tau}$ with $\tau=3/2$. As we decrease $\phi$, $\tau$ begins to increase from $3/2$ and the scaling region gets smaller and smaller suggesting the damage present in the system drives it away from the (pseudo) spinodal. The crosses indicate the data while the dashed line is the best fit to the data which gives $s \sim s^{-\tau}$ with $\tau \approx 2.07$.}
\label{fig:GRvarphi}
\end{center}
\end{figure}

In Fig.~\ref{fig:GRvarphi} we plot the cluster data on a log-log plot for the case in which stress is passed to the dead sites and thus dissipated from the system. 
In the OFC model with long range stress transfer, $n_{s}\sim s^{-3/2}$ which is consistent with Fig.~\ref{sfig:1}. We can see from Fig.~\ref{sfig:09} and~\ref{sfig:07} that the scaling range decreases and eventually disappears as $\phi$ decreases from unity. The question remains as to whether the motion away from the pseudo-spinodal is due to the increased dissipation associated with the dead sites or simply due to the dead sites themselves. We consider this question below. However, there are two interesting points associated with the model as run above. 

\begin{figure}
\begin{center}
\includegraphics[width = \columnwidth]{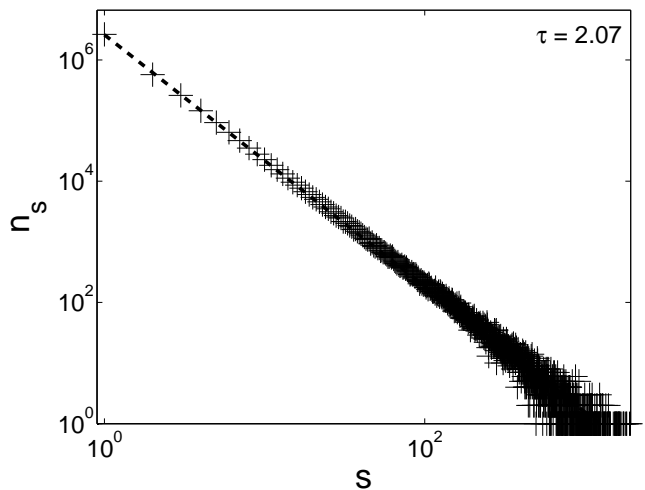}
\caption{The number of clusters ($n_s$) of size $s$ plotted on a log-log plot for all values of $\phi$. We find that there does appear to be a scaling law but the exponent is now approximately $2$. The crosses indicate the data while the dashed line is the best fit to the scaling regime.}
\label{fig:GRallphi}
\end{center}
\end{figure}

\begin{figure}
\begin{center}
\includegraphics[width = \columnwidth]{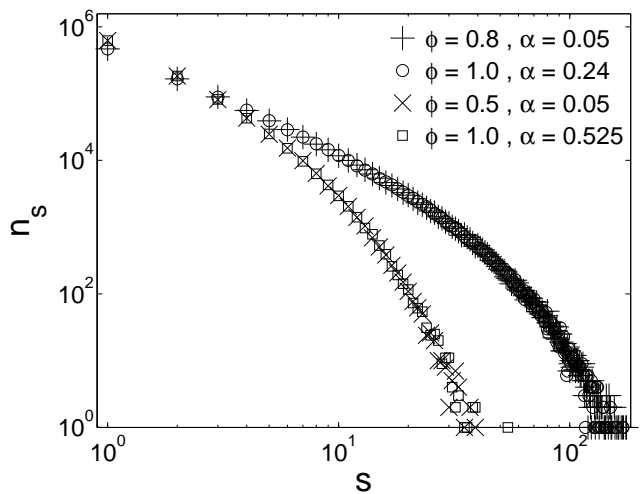}
\caption{The number of clusters ($n_s$) of size $s$ plotted on a log-log plot where stress is passed to all sites for $\phi = 0.8$ and $\phi = 0.5$ (crosses and exes, respectively) and their corresponding undamaged system with $\alpha'=1-\phi(1-\alpha)$ (circles for the system corresponding to $\phi = 0.8$ and boxes for the system corresponding to $\phi = 0.5$).}
\label{fig:GRaeff}
\end{center}
\end{figure}

First, if we consider all of the data generated by the various values of $\phi$ and, again, plot the number of clusters ($n_{s}$) of size $s$, we get what appears to be a scaling law as scene in Fig.~\ref{fig:GRallphi}.
The fact that this would appear to be a scaling plot is due to the fact that the slope at the large cluster end is dominated by the values of $\phi$ near one and the contributions from $\phi \lesssim 0.5$ are concentrated in the region $n_s \lesssim 10$. 
Additionally, we find that running the system with these ``frozen in" dead sites, seems to be the same as running the undamaged model, but with a {\it higher dissipation} parameter. In fact, the lattice described by the damage parameter $\phi$ can be associated with an undamaged system running with a dissipation 
\begin{equation}
\alpha'=1-\phi(1-\alpha)\,, 
\end{equation}
where $\alpha$ is the dissipation parameter for the system being run on a damaged lattice. This is numerically confirmed by noting that the two scaling plots generated by the two different systems with $\alpha$ and $\phi$ related as above lie one on top of the other (see Fig.~\ref{fig:GRaeff}).

\begin{figure}
\begin{center}
\includegraphics[width = \columnwidth]{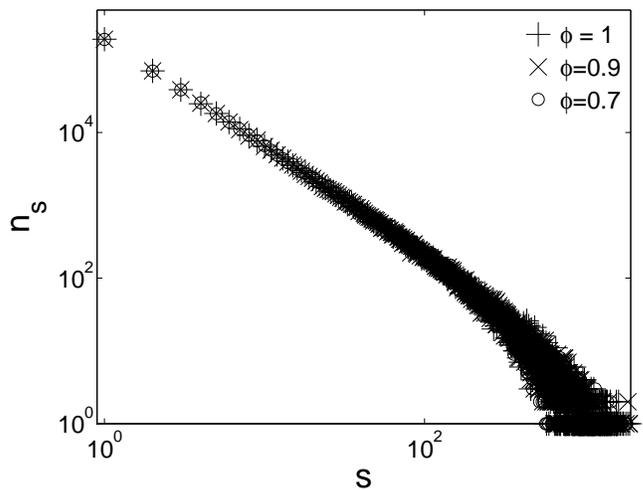}
\caption{The number of clusters ($n_s$) of size $s$ plotted on a log-log plot for various values of $\phi$ where stress is passed only to the live sites. For all values of $\phi$, we get the same scaling behavior as the pure OFC model where $n_s\sim s^{-\tau}$ with $\tau=3/2$.}
\label{fig:GRls}
\end{center}
\end{figure}

\begin{figure}
\begin{center}
\includegraphics[width = \columnwidth]{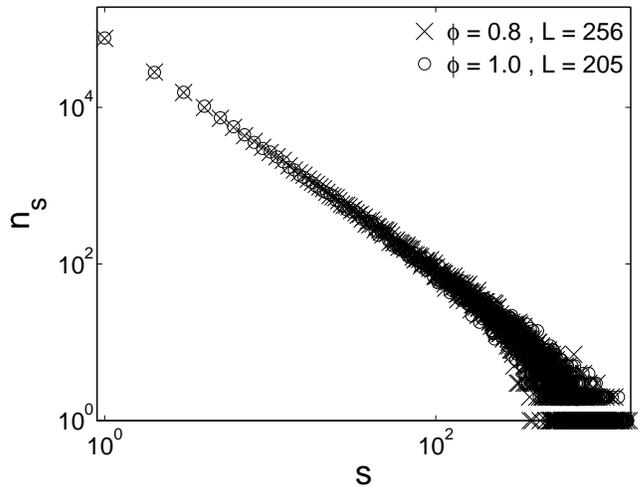}
\caption{The number of clusters ($n_s$) of size $s$ plotted on a log-log plot where stress is passed only to the live sites for $\phi = 0.8$ (exes) and its corresponding undamaged system with $L'=\sqrt{\phi}\,L$ (boxes). Both scale as $n_s\sim s^{-\tau}$ with $\tau=3/2$.}
\label{fig:GRleff}
\end{center}
\end{figure}

The second case we consider is when only {\it live} neighbors and not all of the the neighbors of a failed site evenly share the discharged stress. This is the model considered in Section~\ref{sec:ergodic} where we plot the TM metric for the OFC model with damage. In Fig.~\ref{fig:GRls}, we plot the number of clusters of size $n_s$ versus $s$ when stress is only transferred to live sites.
As can be seen from the figure, when the stress is transferred only to live sites the scaling is the same as the scaling in the pure model. The addition of the damage seems to be equivalent to simulating a smaller system. This can be seen in Fig.~\ref{fig:GRleff} where we compare the scaling plots generated by two systems, one of which has linear dimension $L$ and damage parameter $\phi < 1$, and the other is a non-damaged ($\phi = 1$) system with linear dimension 
\begin{equation}
L' = \sqrt{\phi}\,L\,.
\end{equation}

\section{Long v. Short Range\label{sec:range}}

\begin{figure}
\begin{center}
\subfigure[]{\includegraphics[scale = 0.2]{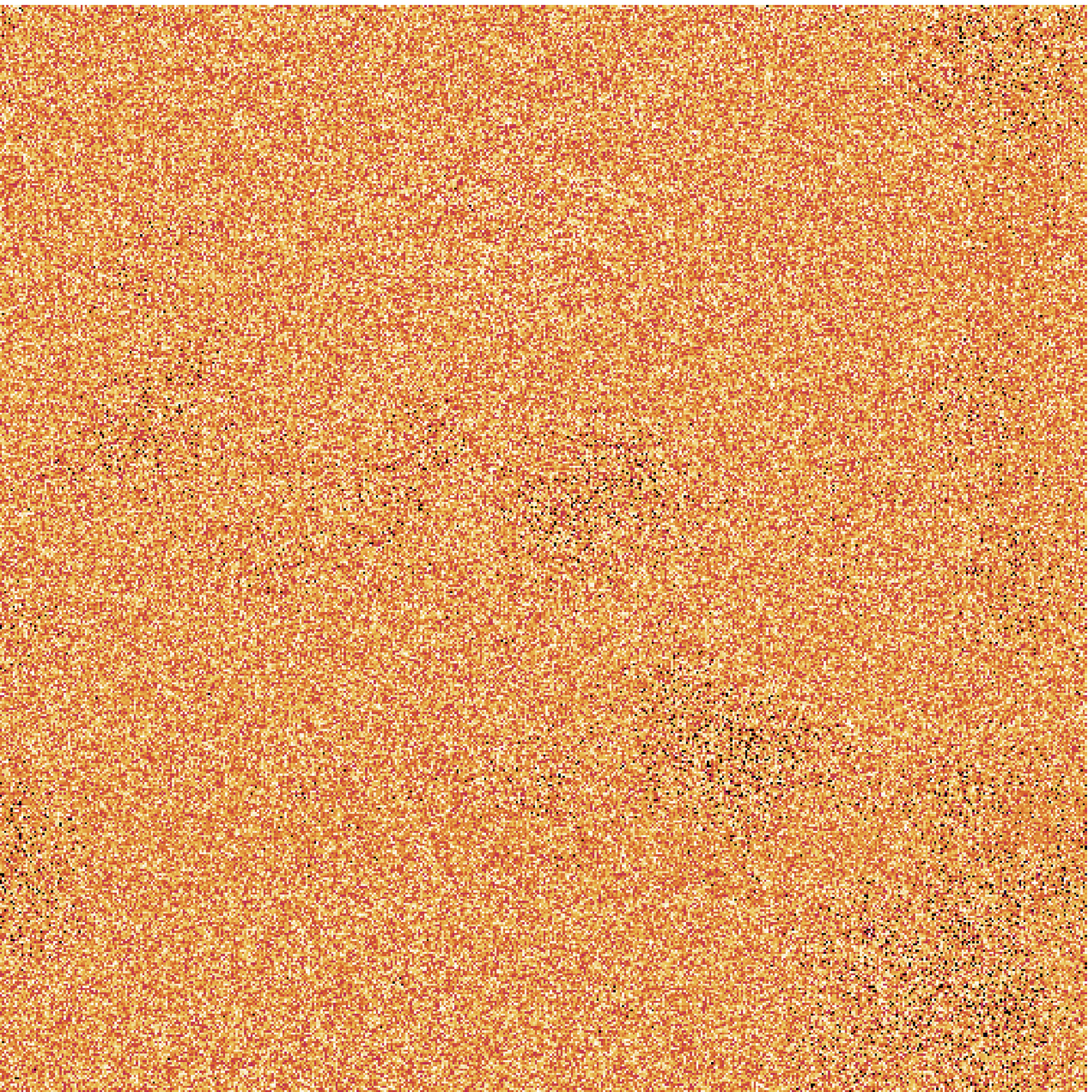}} \hspace{5mm}
\subfigure[]{\includegraphics[scale = 0.2]{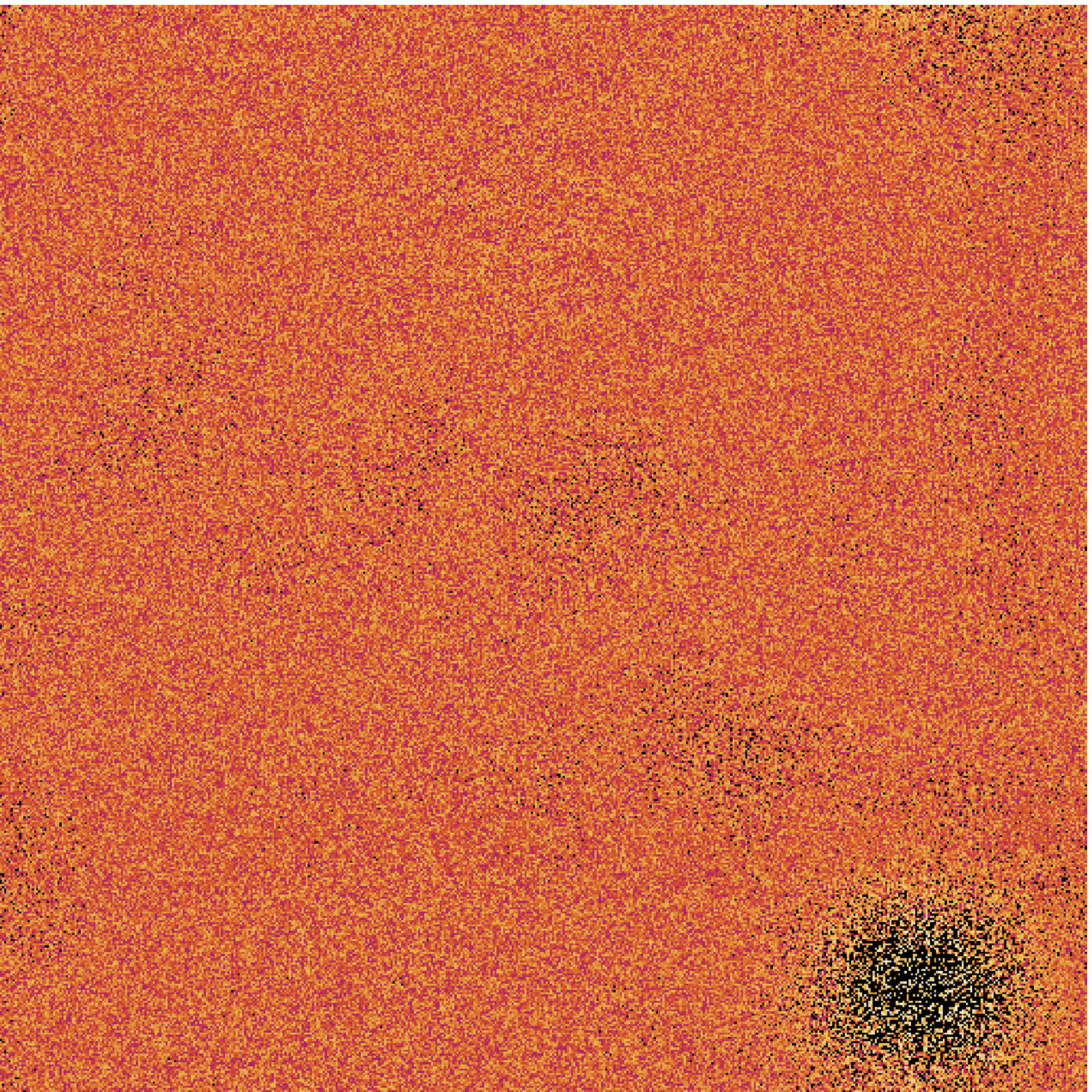}} \\
\subfigure[]{\includegraphics[scale = 0.2]{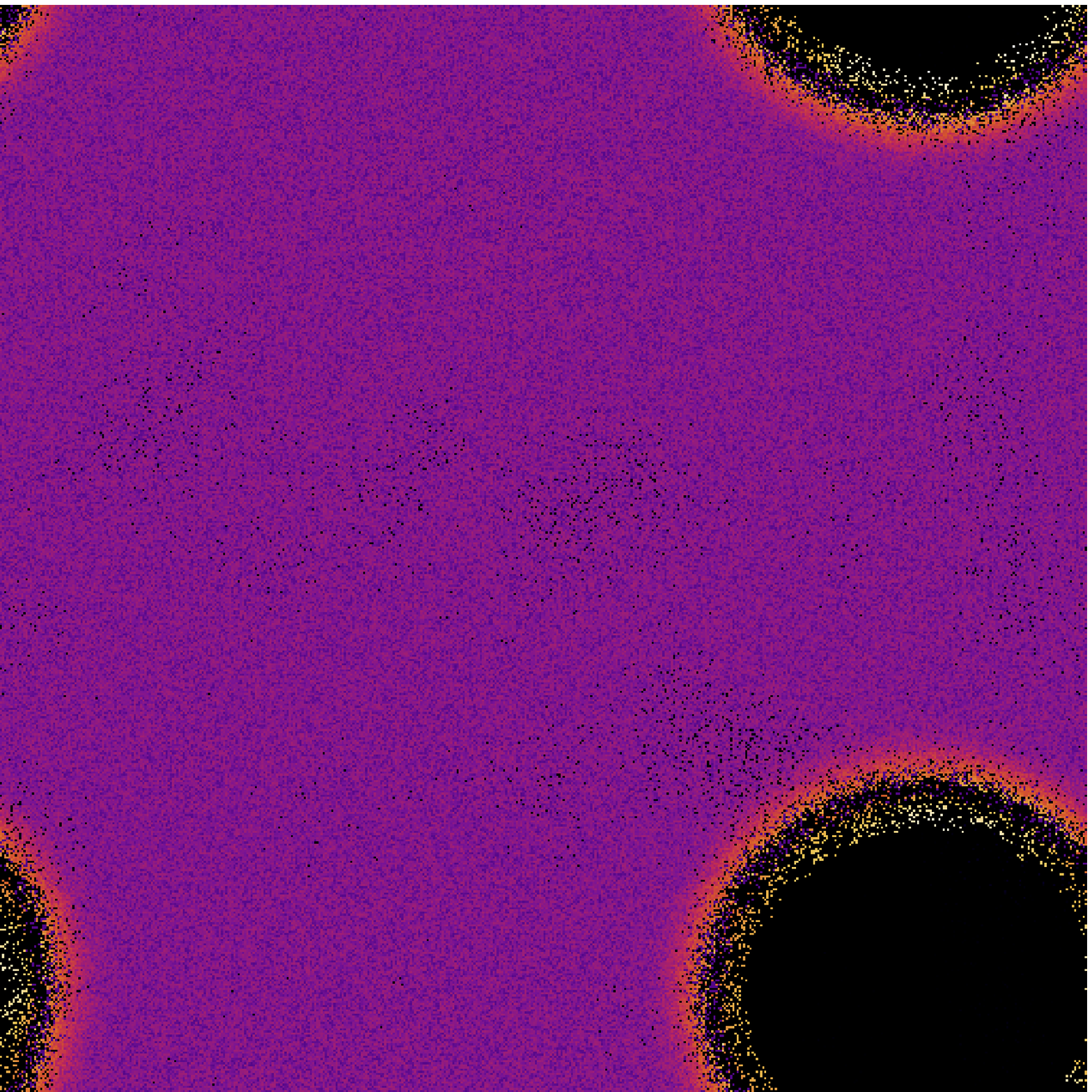}} \hspace{5mm}
\subfigure[]{\includegraphics[scale = 0.2]{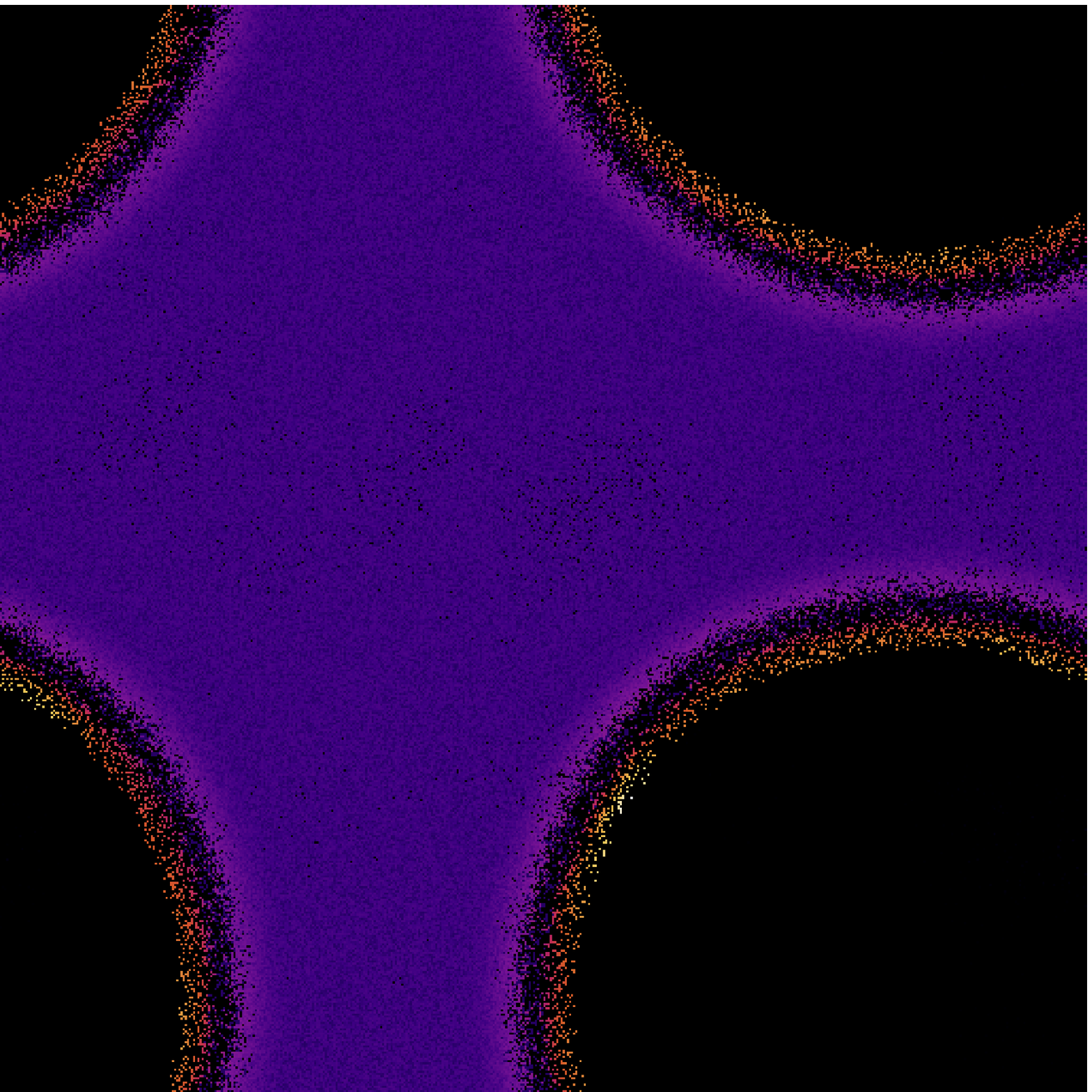}} 
\caption{(Color online) The catastrophic failure event in the long range base model. The event begins in a dense region of dead sites (a) which, locally, overwhelm the system (b) and grow outward (c) in the shape of the interaction, ultimately failing the entire lattice (d) in a single time step.}
\label{fig:LRfail}
\end{center}
\end{figure}

\begin{figure}
\begin{center}
\subfigure[]{\includegraphics[scale = 0.2]{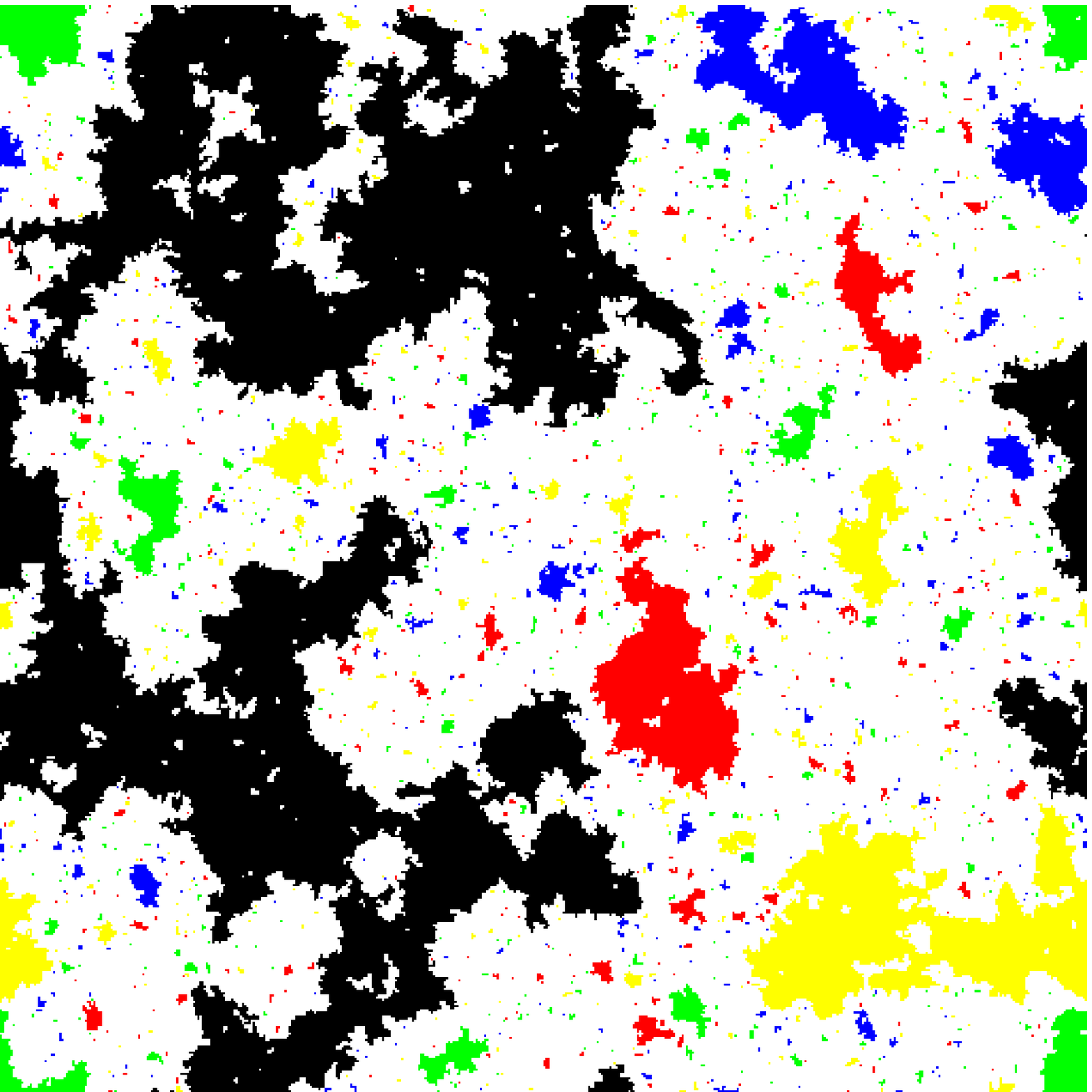}} \hspace{5mm}
\subfigure[]{\includegraphics[scale = 0.2]{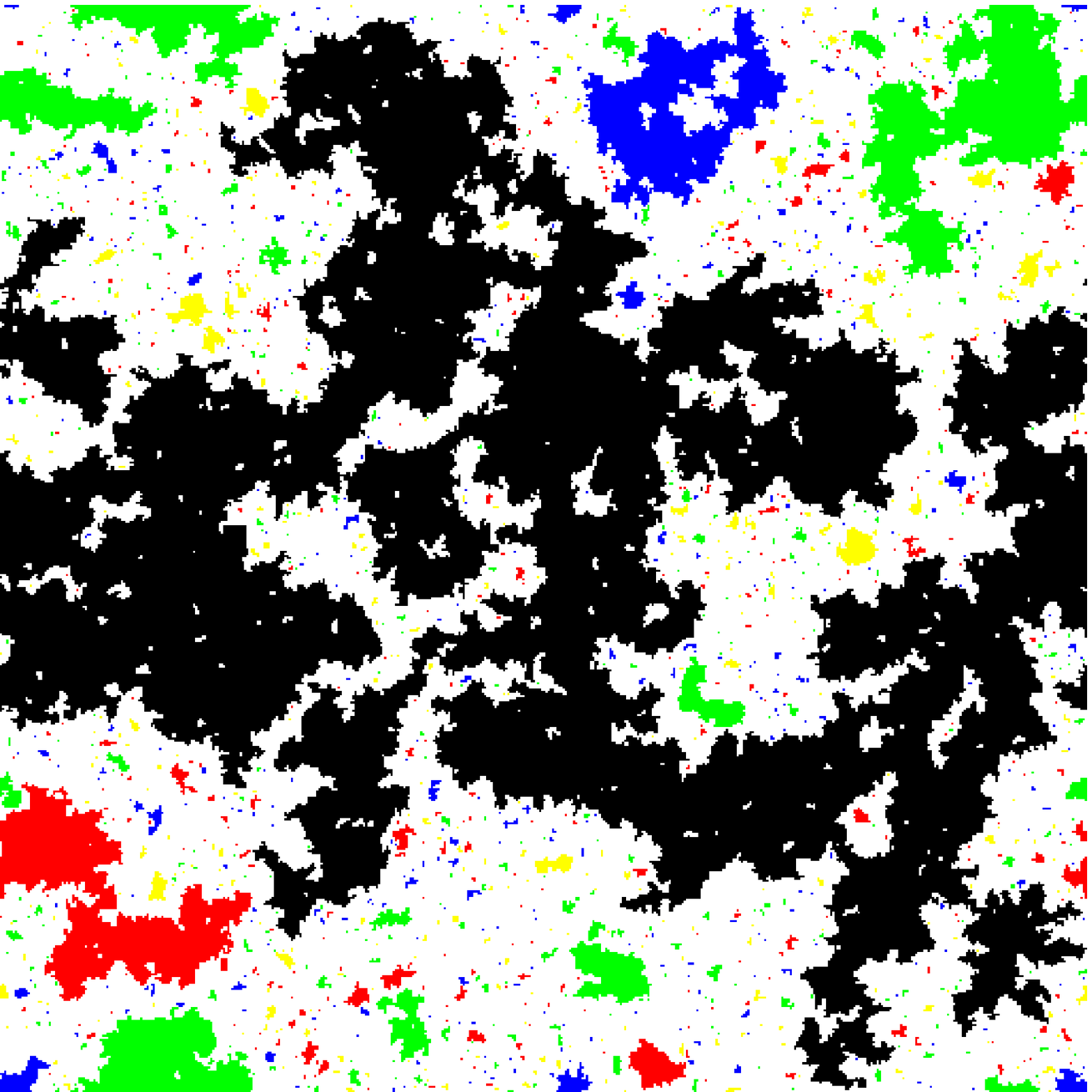}} \\
\subfigure[]{\includegraphics[scale = 0.2]{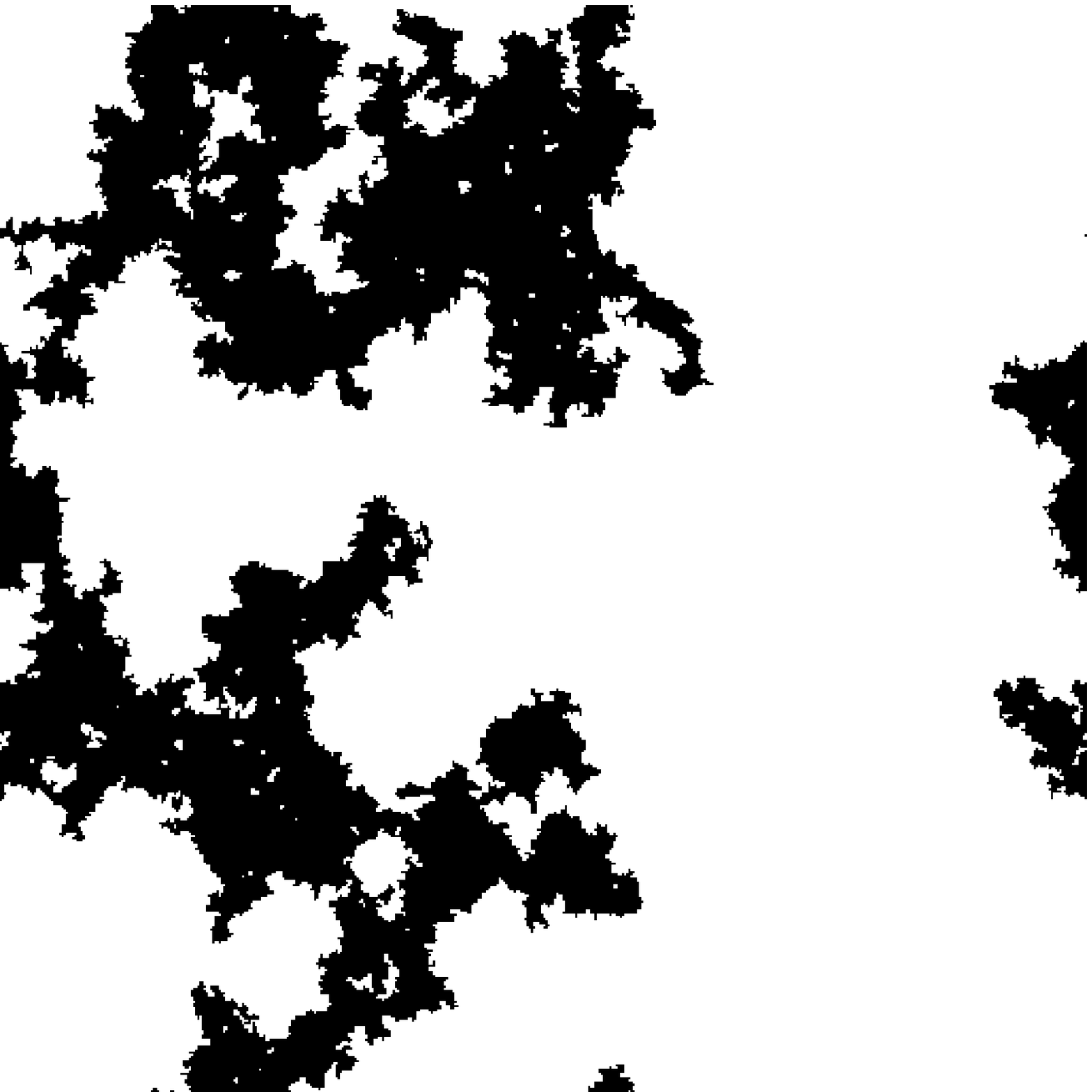}} \hspace{5mm}
\subfigure[]{\includegraphics[scale = 0.2]{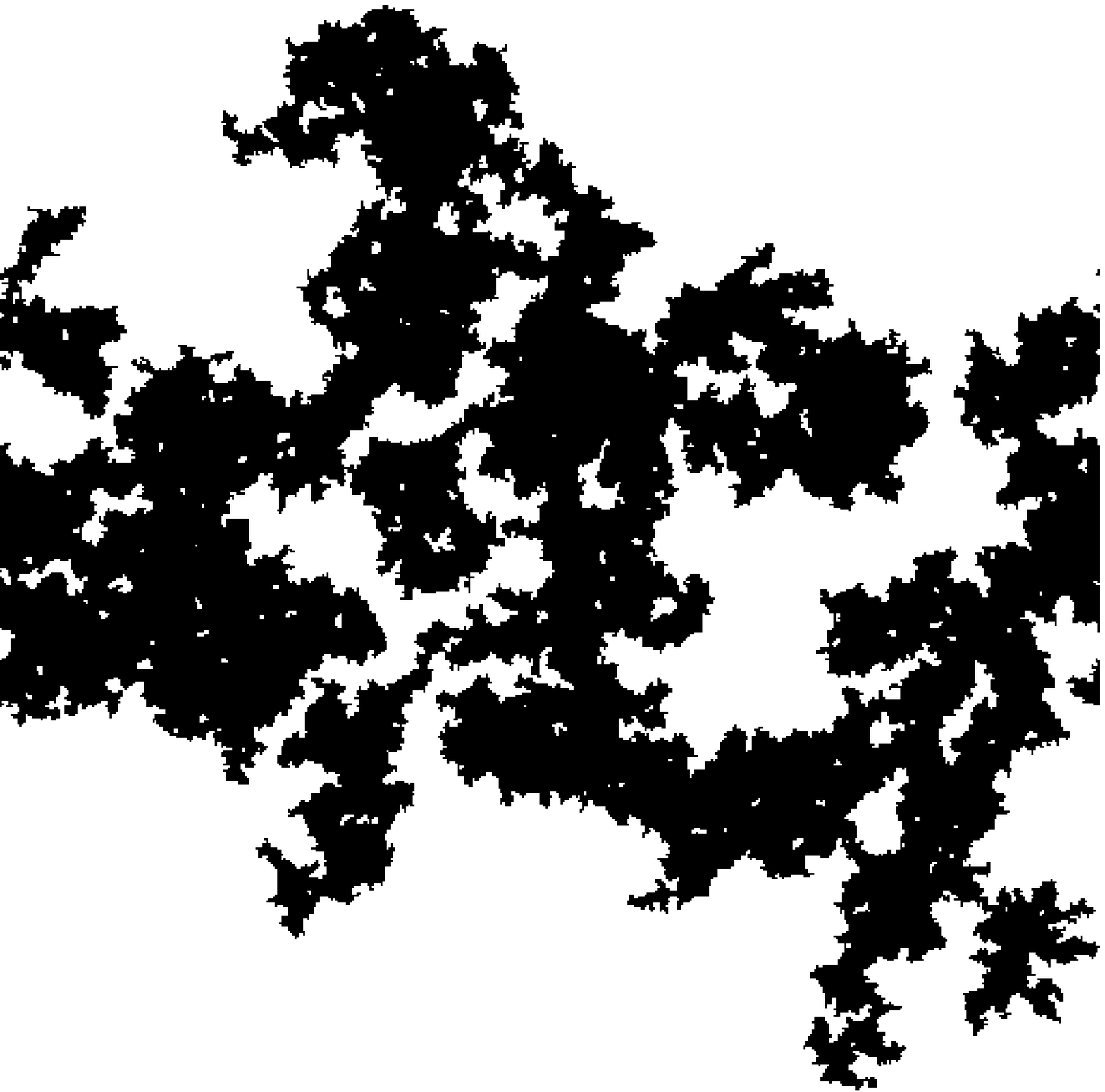}} 
\caption{(Color online) Two typical nearest-neighbor base model lattices at the time of catastrophic failure (percolating cluster). Figures (a) and (b) show all of the clusters at the time the spanning cluster appears. Figures (c) and (d) isolate the percolating cluster from figures (a) and (b), respectively.}
\label{fig:NNfail}
\end{center}
\end{figure}

In this section, we restrict our attention to the base model and study the geometry of the catastrophic failure as a function of the stress transfer range. \footnote{Movies of the catastrophic failure (``nucleation") event similar to Figs.~\ref{fig:LRfail}~and~\ref{fig:NNfail} can be found at \url{http://physics.bu.edu/\~cserino/DamageMovies}.} We find that the geometry of catastrophic failure in the long range stress transfer case is different then in the short range case. In the long range model, when the system undergoes catastrophic failure the lattice goes from about 30\% dead to 100\% dead in one time step (see Fig.~\ref{fig:LRfail}). The process begins in a localized region and appears to be similar to a nucleation event, with a droplet whose interior is the stable phase. In the case of short range stress transfer one has to define catastrophic failure a bit more carefully. If we are using this model to simulate a FBM then catastrophic failure is defined as 100\% dead. This is a gradual process which does not resemble the process in the long range stress transfer model. However, we can also think of this model (as well as the model with long range stress transfer) as a chip board or material such as a rock. In this case, catastrophic failure is defined as a cluster of dead sites that span the system or, in other words, the dead sites form a percolating cluster. Note that this is a different use of the term cluster than in Section~\ref{sec:critpt}. Here, cluster refers to a set of dead lattice sites that are connected to one and other as in nearest-neighbor random site percolation, namely, two nearest neighbor dead sites belong to the same cluster. In this case, the fraction of dead sites is between 30\% and 80\% at the time of catastrophic failure (percolation) and never reaches the state where 100\% of the sites are dead (see Fig.~\ref{fig:NNfail}). Note that the critical percolation density for nearest-neighbor random site percolation in a two dimensional square lattice is approximately 0.593. \cite{aharonystauff}


The short range model has an interesting failure dynamic. As mentioned above, rather than the catastrophic failure occurring suddenly, i.e. in one plate update, the short range model percolates and reaches a mechanical equilibrium (i.e. $\sigma_i < \sigma^f$ for all live sites) before it reaches the point of 100\% failed. The failure begins with small independent clusters of dead sites. As the evolution continues, some of the clusters begin to merge to form larger clusters. Eventually, enough clusters merge so that the system is separated into two pieces, one inaccessible to the other without passing through the spanning cluster. Two typical percolating clusters are shown in Fig.~\ref{fig:NNfail}. The top two figures show the entire lattice with unique clusters corresponding to unique colors and the bottom two figures show only the percolating cluster.
We analyzed the fractal dimension $d_{f}$ of the spanning cluster and found $d_{f}\approx 1.85$ This is consistent with the fractal dimension of two dimensional random percolation, where $d_{f} = 1.896$ \cite{aharonystauff}, within the accuracy of our measurements.

\section{Conclusion\label{sec:conc}}

We have introduced a new model for the study of damage based on the OFC model for earthquake faults. 
The primary change is that we allow sites to die after a proscribed number of failures.  We study two cases: one in which dead sites dissipate all of the stress that is passed to them and one in which dead sites are not allowed to receive stress. 
These dead sites mimic damaged elements such as broken fibers in the fiber bundle models or cracks in fault systems or materials such as chip boards. Our numerical investigation of this model has produced the following results: The system ceases to be ergodic, and hence is not in equilibrium, in the sense described in Section II, as soon as sites begin to die. Healing the sites on a time scale of one plate update or more does not restore the ergodicity seen in the OFC model. The presence of dead sites seems to not only drive the system out of equilibrium but also drives it away from the (pseudo) spinodal in the case of long range stress transfer where the stress is transferred to the dead sites causing additional dissipation. If the stress is only transferred to live sites then the system remains near the (pseudo) spinodal.

Catastrophic failure in the long range system resembles classical nucleation. However, this requires further investigation.  First, the system is not in metastable equilibrium as can be seen from the TM metric data so the standard quasi-equilibrium methods \cite{langer67,rundlekl89} do not apply. Second, the ``nucleation'' process that causes catastrophic failure is not seen in the nearest neighbor stress transfer system so there exists some crossover regime that needs to be studied. Catastrophic failure in the short range stress transfer system does not resemble nucleation but can be classified as a continuous process. In the case where catastrophic failure is defined as the lattice being split into two separated pieces by a percolating cluster of dead sites, analogous to the fracturing of a chip board \cite{chipboard}, the catastrophic failure event can be classified as a fractal. The research presented on this model suggests several directions for further investigations into the nature of damage and catastrophic failure.

\begin{acknowledgments}
This research was supported by the Department of Energy through grant DE-FG02-95ER14498 (CAS and WK) and grant DE-FG03-95ER14499 (JBR).
\end{acknowledgments}


\begin{thebibliography}{damage}

\bibitem{daniels45} H. E. Daniels, Proc. Roy. Soc. London {\bf A 183}, 405 (1945)

\bibitem{peirce26} F. T. Peirce, J. Text. Ind. {\bf 17}, 355 (1926) 

\bibitem{newetal95} W. I. Newman, D. L. Turcotte and A. M. Gabrielov, Phys. Rev. E {\bf 52}, 4287 (1995)

\bibitem{pradhan08} S. Pradhan, A. Hansen and B. K. Chakrabarti, (arXiv 0808.1375, [Cond-Mat.Stat-Mech]) 2008

\bibitem{sornette} D. Sornette J. Phys. A {\bf 22}, L 243 (1989)

\bibitem{herrmann1989} H. J. Herrmann, A. Hansen and S. Roux, Phys. Rev. {\bf B 38}, 637 (1989) 

\bibitem{hidalgo2002} R. C. Hildago, F. Kun and H. J. Herrmann, Phys. Rev. {\bf E 65}, 032502 (2002)

\bibitem{robin91} R.L.B. Selinger, Z. G. Wang, W. M. Gelbart and A. Ben-Shaul, Phys. Rev. {\bf A 43}, 4396 (1991)

\bibitem{sahimi03} M. Sahimi {\it Heterogeneous Materials II: Nonlinear and Breakdown Properties} (Springer-Verlag, Berlin) (2003)

\bibitem{langer67} J. S. Langer, Annals of Physics {\bf 41}, 104 (1967)

\bibitem{rundlekl89} J. B. Rundle and W. Klein, Phys. Rev. Lett. {63}, 171 (1989)

\bibitem{TM90} D. Thirumalai and R. D. Mountain, Phys. Rev {\bf A 42}, 4574 (1990)

\bibitem{DTFB} A. Virgili, A. Petri, and S. R. Salinas, J. Stat. Mech. P04009 (2007), URL http:// stacks.iop.org/1742-5468/2007/P04009.

\bibitem{OFC92} Z. Olami, J. S. Feder and and K. Christensen, Phys. Rev. Lett. {\bf 68}, 1244 (1992)

\bibitem{BK67} R. Burridge and L. Knopoff, Bull. of the Seismol. Soc. {\bf 57}, 341 (1967)

\bibitem{rundle1977} J. B. Rundle and D. D. Jackson, Bull. Seismol. Soc. of AM. {\bf 67}, 1363 (1977)

\bibitem{rundle1991} J. B. Rundle and S. R. Brown, J. Stat. Phys. {\bf 65}, 403 (1991)

\bibitem{rundle1995} J. B. Rundle, W. Klein, S. Gross and D. L. Turcotte, Phys. Rev. Lett. {\bf 75}, 1658 (1995) 

\bibitem{ferguson99} C. F. Ferguson, W. Klein and J. B. Rundle, Phys. Rev. {\bf 60}, 1359 (1999)

\bibitem{klein97} W. Klein, J. B. Rundle and C. F. Ferguson Phys. Rev. Lett. {\bf 78}, 3793 (1997)

\bibitem{klein2000} W. Klein, M. Anghel, C. D. Ferguson, J. B. Rundle and J. S. Sa Martins in {\it GeoComplexity and the Physics of Earthquakes} J. B. Rundle, D. L. Turcotte 
and W. klein eds. (American Geophysical Union, Washington, D. C. 2000) 

\bibitem{Zia} R. P. K. Zia Phys. Rev. Lett. {\bf 66}, 357 (1991)

\bibitem{Schmittmann} B. Schmittmann Eur. Phys. Lett. {\bf 24} 109 (1993)

\bibitem{Praestgaard} E. L. Praestgaard, B. Schmittmann and R. K. P. Zia Eur. Phys. Journ. {\bf B18}, 675 (2000)


\bibitem{klein2007} W. Klein, H. Gould, N. Gulbahce, J. B. Rundle and K. F. Tiampo, Phys. Rev. {\bf E 75}, 031114 (2007)

\bibitem{binder1984} K. Binder, Phys. Rev. {\bf A 29}, 341 (1984)

\bibitem{whiskers} S. S. Brenner, J. Appl. Phys. {\bf 33}, 33 (1962)

\bibitem{rundle1997} J. B. Rundle, W. Klein, S. Gross and D. L. Turcotte, Phys. Rev. Lett. {\bf 78}, 3798 (1997) 

\bibitem{aharonystauff} D. Stauffer and A. Aharony, {\it Introduction to Percolation Theory} Plenum Press, New York (1994)

\bibitem{monette92} L. Monette and W. Klein, Phys. Rev. Lett. {\bf 68}, 2336 (1992)

\bibitem{petri} A. Petri, G. Paparo, A. Vespignani, A. Alippi and M. Costantini, Phys. Rev. Lett. {\bf 73}, 3423 (1994)

\bibitem{chipboard} A. Garcimart\'{\i}n, A. Guarino, L. Bellon and S. Ciliberto, Phys. Rev. Lett. {\bf 79}, 3202 (1997)

\bibitem{note1} The reason that the TFB model with global load sharing cannot nucleate is that the energy is quadratic in the strain and the stress is linear in the strain. Therefore as more fibers fail the remaining fibers must get longer to keep the global stress constant. This generates an increasing energy cost for the same change in length as the fiber gets longer. Hence it becomes energetically impossible to get over the barrier between the meta-stable state with a fraction of the fibers intact and the stable state of all fibers broken as the free energy cost, $\Delta F \sim 1/\phi$ for $\phi \to 0$. The simulations in Selinger \etal\ model a one dimensional crack embedded in a two dimensional medium. This allows the energy dissipated by the broken fibers on the crack to be deposited in a reservoir consisting of the medium away from the crack.


\end{thebibliography}
\end{document}